\documentclass[preprint,aps,12pt]{revtex4}
\usepackage{amsmath}
\usepackage{amssymb}
\usepackage{epsfig}
\usepackage{slashed}
\usepackage{siunitx}
\usepackage{graphicx}
\usepackage{multirow}
\def\missE{\slashed E} 
\begin{document}
\title{ Search for anomalous couplings via single top quark production in association with a photon at LHC }
\author{Yu-Chen Guo $^{1}$}
\author{ Chong-Xing Yue $^{1}$}\email{cxyue@lnnu.edu.cn}
\author{ Shuo  Yang $^{2}$ }

\affiliation{
$^1$ Department of Physics, Liaoning Normal University, Dalian 116029,  China\\
$^2$ Physics Department, Dalian University, Dalian 116622,  China
\vspace*{1.5cm}}

\begin{abstract}
Considering the experimental constraints given by the CMS collaboration at $\sqrt{s} = 8$ TeV on the strength of top quark flavour-changing neutral-current couplings $tq\gamma$ and $tqg$, we investigate the production of top quark in association with a photon and carry out a full simulation for the signals $\ell\nu b\gamma$ and $jjb\gamma$ at 14 TeV LHC. In our numerical analysis, the contributions of single top production with a photon radiation off the top decay products are also included. The discovery potential for anomalous couplings  $tq\gamma$ and $tqg$ with an integrated luminosity of 100 fb$^{-1}$ are examined in detail.

\end{abstract}


\maketitle
\section{Introduction}

The top quark is generally considered as a sensitive probe of physics beyond the standard model (BSM)\cite{Top}, with its large mass close to the electroweak symmetry breaking scale. Transitions between top quarks and other quark flavours mediated by neutral gauge bosons, the flavour-changing neutral-currents (FCNC), are forbidden at tree level in the standard model (SM) and suppressed at the level of quantum loop due to the Glashow-Iliopoulos-Maiani (GIM) mechanism\cite{GIM}. The branching ratios for the processes $t \rightarrow q\gamma$ and $t \rightarrow qg$ are of the order of $10^{-14}-10^{-12}$ in the SM \cite{topBr1,topBr2}. In contrast, several BSM scenarios, such as the two-Higgs doublet model, supersymmetry or technicolor, predict much larger rates \cite{BSMtopBr1,BSMtopBr2} of the order of  $10^{-6}- 10^{-5}$. It implies that observation of the large FCNC-induced couplings $tq\gamma$ and $tqg$ would indicate the existence of BSM.

The enhanced FCNC $tq\gamma$ and $tqg$ interactions are predicted by many extensions of the SM which include new exotic quarks \cite{exotic quarks}, new scalars \cite{new scalars1,new scalars2}, supersymmetry \cite{BSMtopBr1,SS1,SS2,SS3,SS4,Lagrangain2}, or technicolour \cite{BSMtopBr2,0801.0210}.
The BSM effects can be described by a minimal set of the higher order effective operators independently from the underlying theory \cite{Lagrangain}. The effective operators not only simplify multiple free parameters of specific models in a model-independent way, but also they order them and allow us to consistently take into account higher order quantum corrections. This method appears in many studies to search top quark FCNC \cite{9909222,0810.3889,0910.4349,1003.3173,1004.0620,1004.0898,1007.2551,1101.5346,1103.5122,1104.4945,1304.5551,1305.2427,1305.7386,1404.1264,1402.1817,1408.0493,1412.5594,1412.7166,NPB897,1511.00220,1601.08193,1305.7096,1601.04165}. It can facilitate the analysis of new physics effects in $tq\gamma$ and $tqg$ interactions.
So we can give limits on the strength of anomalous top couplings in a model-independent way.
The most general effective Lagrangian can be written as
\begin{eqnarray}
\label{lagrangy}
\mathcal{L}_\mathrm{eff} = &-&e Q_{t}\bar{u}\frac{i\sigma^{\mu\nu}q_{\nu}}{\Lambda} (\kappa_{tu\gamma}^{L}P_{L}+\kappa_{tu\gamma}^{R} P_{R})tA_{\mu}
\nonumber\\
&-& e Q_{t}\bar{c}\frac{i\sigma^{\mu\nu}q_{\nu}}{\Lambda} (\kappa_{tc\gamma}^{L}P_{L}+\kappa_{tc\gamma}^{R} P_{R})tA_{\mu} \nonumber\\
&-& g_{s}\bar{u}\frac{i\sigma^{\mu\nu}q_{\nu}}{\Lambda} (\kappa_{tug}^{L} P_{L}+\kappa_{tug}^{R} P_{R})T^a tG_{a\mu}  \nonumber\\
&-& g_{s}\bar{c}\frac{i\sigma^{\mu\nu}q_{\nu}}{\Lambda} (\kappa_{tcg}^{L} P_{L}+\kappa_{tcg}^{R} P_{R})T^a tG_{a\mu} + h.c.,\quad,
\label{trcrosssection}
\end{eqnarray}
where $Q_{t}$ is the electric charge of the top quark, $g_{s}$ is the strong-coupling constant, $T^{a}=\lambda_{a}/2$ are colour matrices, $q$ is the momentum of the gauge boson, and $P_{L(R)}$ denotes the left(right)-handed projection operators. ${\Lambda}$ is the new physics scale, which is related to the cutoff mass scale above which the effective theory breaks down. The terms with $\sigma^{\mu\nu}=\frac{1}{2}[\gamma^{\mu},\gamma^{\nu}]$ are suppressed by the GIM mechanism, and in consequence are absent at tree level in renormalizable theories, like the SM.
Real dimensionless parameters $\kappa_{tqV}^{L,R}$ are the strength of anomalous couplings $tqV$ with $V=\gamma,g$ and $q=u,c$.

Among FCNC top quark decays, $t \rightarrow qg$ is very difficult to distinguish from generic multijet production via
quantum chromodynamics (QCD). It has therefore been suggested to search for FCNC couplings in anomalous single top quark production. The existence of anomalous couplings $tq\gamma$ and $tqg$ would induce production of a top quark in association with a photon, $pp\rightarrow t\gamma$.
Next-to-leading order QCD predictions for this process have been studied in \cite{1101.5346,1412.5594}. This process has been probed at the CMS experiment, as yet, with no indication of any signal.
The strong bounds on the strengths of anomalous couplings have been provided by the CMS experiment with $\sqrt{s}=8$ TeV \cite{CMS003,CMS007}. We consider the CMS limits and give the discovery potential of 14 TeV LHC.

The aim of this paper is to investigate the limits on anomalous top couplings by considering $t\gamma$ production. Unlike previous studies of $t\gamma$ production, we focus on an analysis of signal and backgrounds based on the CMS detector simulation. In addition, we discuss the sensitivity of 14 TeV LHC to anomalous top couplings and detection potential bounds on the $tq\gamma$ and $tqg$ couplings.

The rest of the paper is organized as follows. In Sect. II, we provide the cross sections of $t\gamma$ production with $\sqrt{s}=14$ TeV.
The simulation of signal and the expected backgrounds is discussed in detail. At the end of this section, we discuss the contribution of single top production with a photon radiation off the top decay products.
In Sect. III, we analyze the sensitivity of 14 TeV LHC to anomalous top couplings in detail. The limits on the branching ratios of top quarks into lighter quarks and photons or gluons are given correspondingly.
Finally we summarize our results in Sect. IV.

\section{Anomalous top couplings and $t\gamma$ production at LHC }

In this section we first study $t\gamma$ production with the on-shell top quark in the final state, and then discuss the signal and background events for two channels $jjb\gamma$ and $\ell\nu b\gamma$ depending on the $W$ decay mode. A realistic and detailed analysis is presented, including object identification and event selection.
\begin{figure}[ht]
\begin{center}
\includegraphics [scale=1] {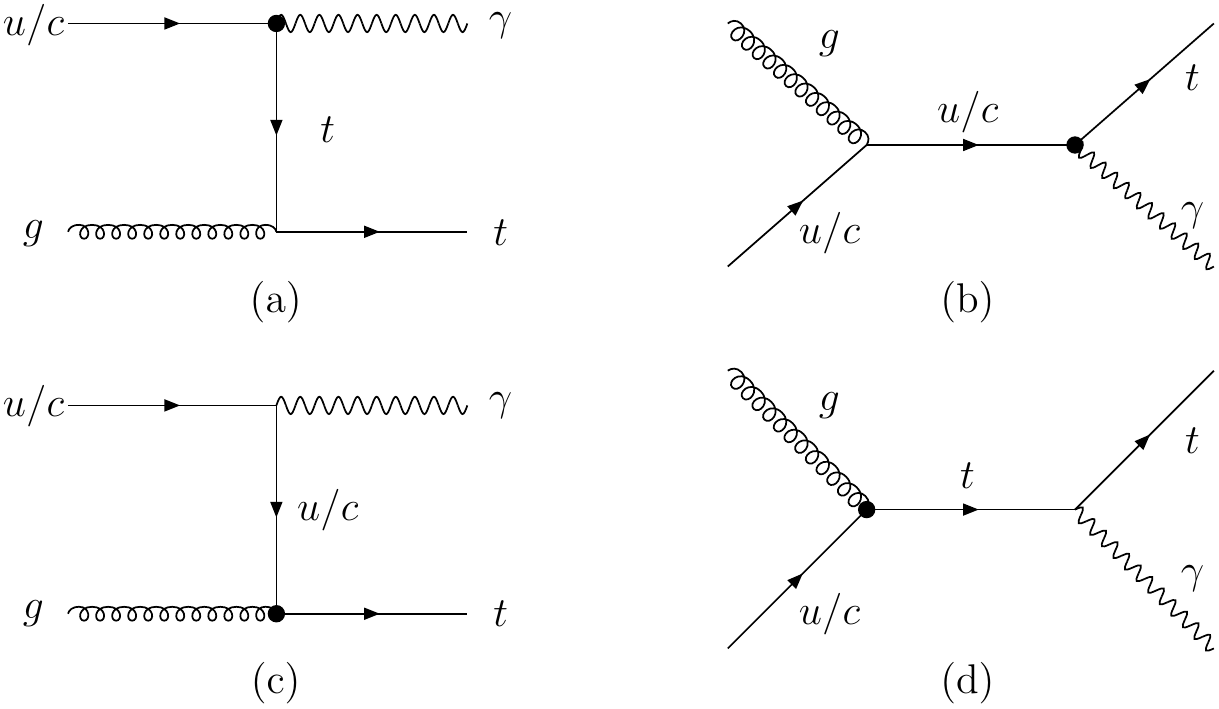}
\caption{Leading order Feynman diagrams for anomalous $t\gamma$ production.
Flavor violation (the dots) occurs either in the weak (a,b) or strong (c,d) sector. } \label{Feynman}
\end{center}
\end{figure}

\subsection{$t\gamma$ production}
From the general effective Lagrangian $\mathcal{L}_\mathrm{eff}$ as shown in Eq. (\ref{lagrangy}), one can see that anomalous couplings $tq\gamma$ and $tqg$ induce the process $pp\rightarrow t\gamma$. The $t\gamma$ production provides the opportunity to probe both anomalous $tuV$ and $tcV$ couplings. Less sensitivity to $tcV$ couplings is expected due to the large enhancement of the charm quark parton distribution function (PDF). Therefore, we consider that each type of interaction, $tu\gamma$, $tc\gamma$ and $tug$, $tcg$, should be treated independently.
In our analysis, we take ${\Lambda}$ as the top quark mass, $m_{t}=173.2$ GeV, $\alpha_s= 0.108$, $\alpha= 1/128.92$ and a simplified scenario with $\kappa_{tq\gamma}^{L}=\kappa_{tq\gamma}^{R}=\kappa_{tq\gamma}$ and $\kappa_{tqg}^{L}=\kappa_{tqg}^{R}=\kappa_{tqg}$.

The presence of the anomalous couplings $tq\gamma$ and $tqg$ leads to the single top production in association with an energetic photon by two main mechanisms related to the strong and weak sector.
This process can take place through the s- and t-channel. The corresponding Feynman diagrams are shown in Fig. \ref{Feynman}.
The effective cross sections $\sigma(s)$ can be evaluated from $\hat{\sigma}(\hat{s})$ by
convoluting with $f_{q_1/p}(x_1)$ and $f_{q_2/p}(x_2)$,
\begin{eqnarray}
\sigma(s)=\int^1_{x_\mathrm{min}} \mathrm{d}x_1 \int^1_{x_\mathrm{min}/x_1} \mathrm{d}x_2 f_{q_1/p}(x_1) f_{q_2/p}(x_2) \hat{\sigma}(\hat{s}),
\end{eqnarray}
where $\hat{s}=x_1x_2s$ is the effective center-of-mass (c. m.) energy squared for the partonic process,
and $x_\mathrm{min}=m_{t}^2/s$.
For the quark distribution functions $f_{q_1/p}(x_1)$ and $f_{q_2/p}(x_2)$, we will use the form
given by the leading order parton distribution function CT14 \cite{CTEQ}.

The effective Lagrangian is implemented in FeynRules \cite{Feynrules} and subsequently passed to Madgraph5/aMC@NLO\cite{mg5amc} framework by means of UFO module \cite{UFO}. We assume that only one anomalous top coupling is nonzero. It is worth mentioning that if the photon is collinear to the initial quark, the cross section will have a divergence. To avoid this divergence, we set a minimum transverse momentum cut on emitted photons, $p_{T}^{\gamma} > 50$~GeV which is adopted by the CMS collaboration.  In this case, the $t\gamma$ cross section at 14 TeV are

\begin{equation}
\begin{split}
  \sigma_{t\gamma}(\kappa_{tu\gamma}) =&\  \phantom{2}144.4\ \big|\kappa_{tu\gamma}|^2\ (pb) \ , \\
  \sigma_{t\gamma}(\kappa_{tc\gamma}) =&\  \quad 13.7\ \big|\kappa_{tc\gamma}|^2\ (pb) \ , \\
  \sigma_{t\gamma}(\kappa_{tug}) =&\   \phantom{1}401.6\ \big|\kappa_{tug}|^2\ (pb) \ , \\
  \sigma_{t\gamma}(\kappa_{tcg}) =&\   \quad 55.7\ \big|\kappa_{tcg}|^2\ (pb) \ .\\
\end{split}\label{CrossSectionEq}
\end{equation}
Obviously, the cross sections of $t\gamma$ production only depend on the strengths of anomalous top couplings $\kappa_{tq\gamma}$ and $\kappa_{tqg}$.

There are many alternatives for normalisation of coupling constants in $\mathcal{L}_\mathrm{eff}$.
The experimental results always use branching ratios. In order to compare with them, we will show our results by using branching ratios of top quark. In our analysis, the width of $t\rightarrow Wb$ is assumed to be approximately top quark total width. The LO prediction for decay width of top quark decay to a bottom quark and a $W$ boson is \cite{t2wb}
\begin{equation}
\Gamma (t \rightarrow Wb) = \frac{\alpha}{16 s_w^2} |V_{tb}|^2 \frac{m_t^3}{m_W^2} [1-3\frac{m_t^4}{m_W^4}+2\frac{m_t^6}{m_W^6}].
\end{equation}
The partial widths of the top FCNC decays $t\rightarrow q\gamma$ and $t\rightarrow qg$ are expressed as follows:
\begin{eqnarray}
&&\Gamma (t \rightarrow q\gamma) = \frac{2 \alpha}{9} m_t^3 \frac{|\kappa_{q\gamma}|^2}{\Lambda^2}, \nonumber\\
&&\Gamma (t \rightarrow qg) = \frac{2 \alpha_s}{3} m_t^3 \frac{|\kappa_{qg}|^2}{\Lambda^2}.
\end{eqnarray}

\begin{figure}[!htb]
\begin{center}
\includegraphics [scale=0.5] {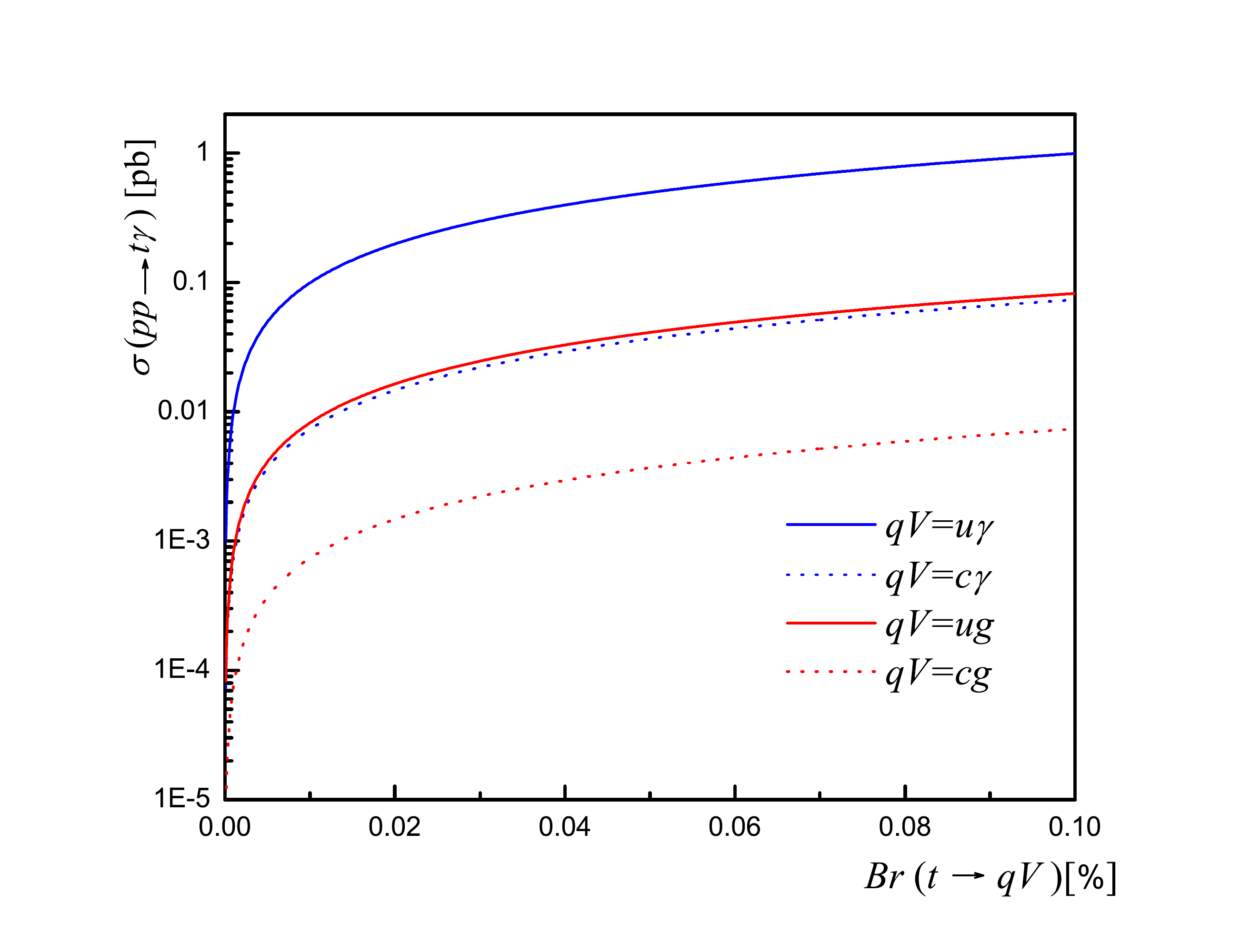}
\vspace{-0.7cm}
\caption{The cross sections of anomalous $t\gamma$ production versus the FCNC branching ratios $\mathcal{B}\mathcal{R}(t\rightarrow u\gamma)$, $\mathcal{B}\mathcal{R}(t\rightarrow c\gamma)$, $\mathcal{B}\mathcal{R}(t\rightarrow ug)$ and $\mathcal{B}\mathcal{R}(t\rightarrow cg)$.} \label{trCrossSection}
\end{center}
\end{figure}

We plot the cross section of $t\gamma$ production originating from different anomalous couplings $tqV$ versus the FCNC branching ratios in Fig.\ref{trCrossSection}.
From Fig.\ref{trCrossSection}, we can see that the cross sections of anomalous $t\gamma$ production increase with the FCNC branching ratios increasing. Experimental limits on the branching ratios of the rare top quark decays were established by experiments of top production at the LEP, HERA, Tevatron and LHC accelerators \cite{CDF1998,CDF2008,D0,ATLAS063,ATLAS2,CMS021,CMS003,CMS007}. At present the most stringent upper limits at $95\%$ confidence level (CL) from different sensitive channels are shown in Table \ref{result}.
Using the upper limits on the branching ratios by the CMS experiment,
\begin{eqnarray}
&&\mathcal{B}\mathcal{R}(t\rightarrow u\gamma) < 1.61\times10^{-4},\
\mathcal{B}\mathcal{R}(t\rightarrow c\gamma) < 1.82\times10^{-3},\ \\
&&\mathcal{B}\mathcal{R}(t\rightarrow ug) < 3.55\times10^{-4},\
\mathcal{B}\mathcal{R}(t\rightarrow cg) < 3.44\times10^{-3},
\end{eqnarray}
we obtain the limits on the cross section in Fig.\ref{trCrossSection}. Then we use Eq.\ref{CrossSectionEq} to calculate the strengths of anomalous top couplings
\begin{eqnarray}
\kappa_{tu\gamma} < 0.028,\ \kappa_{tc\gamma} < 0.094,\ \kappa_{tug} < 0.036 ,\ \kappa_{tcg} < 0.112,
\end{eqnarray}\label{limit}
which are coincident with that given by \cite{CMS003,CMS007}.

\begin{table}[Ht]
\caption{\label{summaryreport} The most stringent experimental upper bounds on the top quark FCNC branching ratios at 95\% CL obtained in CDF, D0, ATLAS and CMS from different channels.}
\begin{tabular}{lllclll}
\hline\hline
EXP & ~$\sqrt{s}$~TeV ~& $~\mathcal{L}(\rm{fb}^{-1})$~&~${\cal B\mathcal{R}}$  &~ $(q=u)\%$ ~&~ $(q=c)\%$~ &~ Ref\\
\hline
CDF &~~~1.8  & ~~~0.11 &\parbox[t]{1mm}{\multirow{3}{*}{\rotatebox[origin=c]{90}{$t\rightarrow q\gamma $ }}}  & \multicolumn{2}{c}{3.2 }  & ~~\cite{CDF1998} \\
CMS &~~~8 & ~~~19.1 &  &~~~~0.0161  & ~~~~0.182 & ~~\cite{CMS003} \\
&&&&&\\
\hline
CDF &~~~1.96   & ~~~~2.2  & \parbox[t]{1mm}{\multirow{5}{*}{\rotatebox[origin=c]{90}{$t\rightarrow qg$ }}}   &~~~~~0.039  & ~~~~~0.57 & ~~\cite{CDF2008} \\
D0 &~~~1.96    & ~~~~2.3  & &~~~~~~0.02  & ~~~~~0.39 & ~~\cite{D0} \\
CMS &~~~7  & ~~~~4.9 & &~~~~~~0.56  & ~~~~~7.12& ~~\cite{CMS021} \\
CMS &~~~7  & ~~~~4.9 & &~~~~~0.035  & ~~~~~0.34 & ~~\cite{CMS007} \\
ATLAS &~~~8  & ~~~14.2  & &~~~~0.0031  & ~~~~0.016 & ~~\cite{ATLAS063} \\
\hline\hline
\end{tabular}\label{result}
\end{table}

\subsection{Signal and background simulation}

Different decay channels of the gauge boson $W$ give different experimental signals. There are two kinds of signals, $\ell \nu b\gamma$ and $jjb\gamma$. We have
\begin{eqnarray}
pp \rightarrow t \gamma \rightarrow  W^+ b \gamma \rightarrow \ell\nu b\gamma ,
\end{eqnarray}
and
\begin{eqnarray}
pp \rightarrow t \gamma \rightarrow  W^+ b \gamma \rightarrow jjb\gamma .
\end{eqnarray}

The leptonic mode of $t\gamma$ production is in general characterized by the presence of an isolated charged leptons (electrons or muons) together with a photon, missing transverse energy, and one $b$-jet.
The $\ell \nu b \gamma$ final state is more attractive from the experimental point of view.
On the one hand, it is relatively efficient to be searched by experiments. Everything in the final state could be the targeted objects. They can be reconstructed efficiently by subdetector systems of LHC detectors. The lepton reconstruction efficiency with $P_t > 5$ GeV is more than $90\%$ and particle identification can be made at detector level. The average efficiency of single photon reaches 91\%. The $b$-tagging algorithm could be used with the efficiency of about $70\%$.
On the other hand, the relatively clean signal is robust against the contamination of pileups and underlying events, since primary collision vertices of the signal events can be reconstructed.
Moreover, $\sigma(\ell^{+})/\sigma(\ell^{-})$ can be used to determine whether $t\gamma$ production comes from the up quark or charm quark initiated process \cite{1104.4945,1402.3073}. It provides the opportunity to understand the underlying new physics.

We use Madgraph5/aMC@NLO to generate signal and backgrounds events in a collision energy $\sqrt{s} = 14$ TeV.
Higher order correction is taken into account for signal by K factor ($K={\sigma_{NLO}}/{\sigma_{LO}}$), which is equal to 1.8 \cite{1101.5346}. For numerical estimation, we take coupling constants $\kappa_{tu\gamma}=0.01$, $\kappa_{tc\gamma}=0.02$, $\kappa_{tug}=0.01$ and $\kappa_{tcg}=0.03$.
Parton showering and fast detector simulations are subsequently performed by PYTHIA6 \cite{py6} and Delphes3 \cite{Delphes}. Jets are clustered by using the anti-$k_{t}$ algorithm with a cone radius $\Delta R = 0.7$ \cite{jet}.

As demonstrated in \cite{CMS003,CMS007}, the signal of FCNC $t\gamma$ production might suffer more realistic experimental issues of fake photon and mis-tagged $b$ jet. The probability for jets to reconstruct a single photon candidate in the electromagnetic calorimeter is about 0.1\%~\cite{CMSTDR}, and the misidentification probability of light quarks or gluons as $b$-jets is approximately $1.5\%$ \cite{CMS2013}.
Thus the light jet could be misidentified as $b$-jet (or photon) candidate and therefore $W\gamma$+jets events will be one of backgrounds with a fake $b$-jet. Similarly, $W$+jets will contribute to backgrounds if two jets are misidentified as an isolated photon and a $b$-jet simultaneously, respectively. For the background events with more than one $b$-jet, we take into account $t\bar{t}$ and $t\bar{t}+\gamma$ and three single-top processes with an additional photon.
The measurement accuracy of the hadronic calorimeter is not enough to distinguish the $W$ or $Z$ boson.
Thereby, $Z\gamma+$jets process also contributes to backgrounds.
For the fully hadronic final state, the overwhelming QCD multijet backgrounds are large.
We consider the main QCD backgrounds $4j$ and $bjjj$ for the signal $jjb\gamma$.

The background processes can be roughly categorized into the following three types:
\begin{itemize}
\item 1) The multijets background processes include $W$+jets, $W\gamma$+jets, $Z\gamma+$jets. To include the QCD effects, we generate multijet events with up to two jets (three for $W$+jets) that are matched to the parton shower using the MLM-scheme \cite{MLM} with merging scale $x_{q} = 15$ GeV.
\item 2) The top processes include $t\bar{t}$, $t\bar{t}+\gamma$ and three single-top production with a photon processes. The reconstructed top mass distribution of signal would be similar to this type of background. The multijets and top background processes contribute to both leptonic and hadronic mode.
\item 3) The QCD processes include $4j$ and $bjjj$. They only affect the $jjb\gamma$ final state.
\end{itemize}
The cross sections of these processes are listed in Table \ref{evgn}. It is worthy of remarking that $W$+jets in the categories above is the dominant background for the leptonic mode, which can greatly affect the significance.

\begin{center}
\begin{table}
  \begin{center}
\begin{tabular}{|c|c|c|c|}
\hline
 & $\sigma$ & Expected number of events  & Number of events \\
 & ( fb ) & at \SI{100}{fb^{-1}}&  generated \\
\hline \hline
$t\gamma$ & \SI{5.234e2}{} & \SI{5.2e4}{} & 100,000\\
\hline \hline
$W$+jets & \SI{3.066e7}{} & \SI{3.1e9}{} & 8,000,000 \\

$W\gamma$+jets & \SI{1.1e5}{} & \SI{1.1e7}{} &  1,000,000 \\

$Z\gamma$+jets & \SI{7.44e4}{} & \SI{7.4e6}{} & 1,000,000  \\
\hline \hline
$t\bar{t}$ & \SI{5.969e5}{} & \SI{6.0e7}{} & 4,000,000  \\

$t\bar{t}\gamma$ & \SI{2.447e3}{} & \SI{2.5e5}{} & 500,000 \\

Single top+$\gamma$ & \SI{1.705e3}{} & \SI{1.7e5}{} & 400,000 \\

\hline \hline
4$j$(QCD) & \SI{2.058e10}{} & \SI{2.06e12}{} & 5,000,000 \\

$bjjj$(QCD) & \SI{2.217e8}{} & \SI{2.2e10}{} & 4,000,000 \\
\hline
\end{tabular}
\end{center}
\caption{\label{evgn}The expected number of events with 100 fb$^{-1}$ integrated luminosity at $\sqrt{s}=14$ TeV and the generated events for all processes are displayed.}
\end{table}
\end{center}

In order to select the most relevant events, we introduce the following preselection cuts:
\begin{itemize}
\item To trigger the signal events, every event is required to have one isolated photon and one $b$-jet. Additionally, $N(\ell)=1$ is applied by the leptonic mode of signal, and $N(j)<4$ for hadronic mode. The number of targeted objects in each events can help to suppress the background events effectively, especially to the events with fake particles.
\item One of the distinctive signatures of the signal is the presence of a high-$p_{T}$ photon in the final state. The photon is expected to carry large momentum because of the recoil against the heavy top quark. Photon candidates with significant energy are required to have transverse momentum $p_T\!\geq\!50$~GeV with $|\eta|\!\leq\!2.5$, using the CMS coordinate system presented~\cite{CMS2007}. Additionally, only leading jet with $p_{T}>30$ GeV for hadronic mode and $p_{t}(\ell)>20$ GeV with $\missE>$30 GeV for leptonic mode are considered in our analysis (Table III).
\item The particle flow isolation $\Delta R=\sqrt{(\Delta \eta)^2+(\Delta \phi)^2}<0.4$ around the photon candidate is applied, where $\Delta\eta$ is the rapidity gap and $\Delta\phi$ is the azimuthal angle gap between the particle pair.  These cuts on photon ensure the events with exactly one photon candidate. In order to have well separated physical objects and remove radiated photons from high $p_{T}$ leptons or final state partons, it is required that $\Delta R$(jet,$\gamma)>0.7$ and $\Delta R($lepton,$\gamma)>0.7$.
\end{itemize}

\begin{center}
\begin{table}
  \begin{center}
  \begin{tabular}{|c|c|}
  \hline
Preselection Cuts & Description  \\ \hline
1 &  $N(\gamma) = 1$ and $N(b)=1$ \\
 & $N(\ell) = 1$ or $N(j)<4$ \\
     \hline \hline
2 & $p_t(\gamma)>50$ GeV,  $p_t(b/j)>30$ GeV,  \\
  & $|\eta|\!\leq\!2.5$, $\Delta R(\ell/j,\gamma)>0.7$, \\
  & $p_t(\ell)>10$ GeV , $\missE>$30 GeV. \\
  \hline
  \end{tabular}
  \end{center}
  \label{precuts}
  \caption{The preselection cuts in our analysis are tabulated.}
\end{table}
\end{center}

\begin{figure}[!htb]
\begin{center}
\includegraphics [scale=0.38] {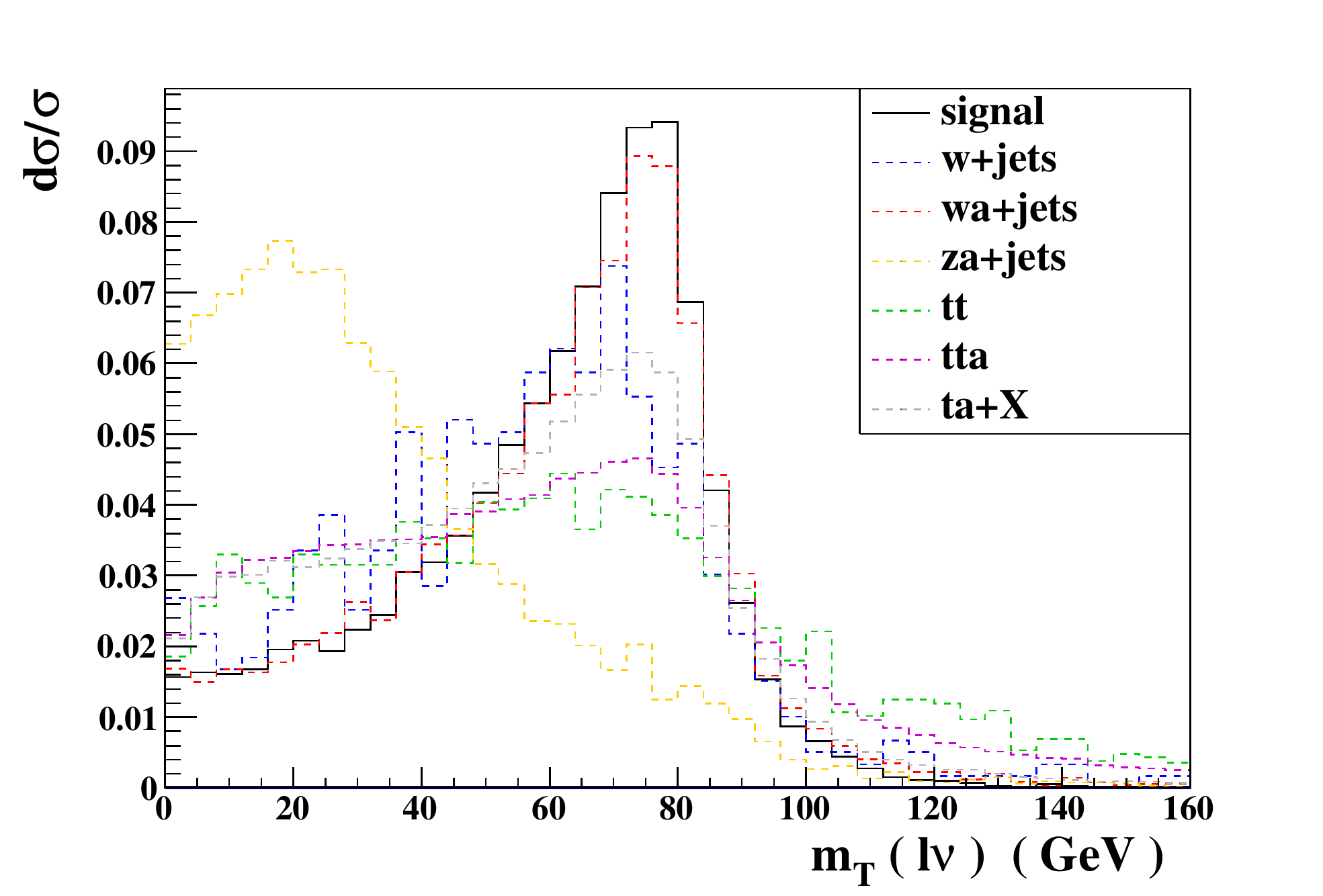}
\includegraphics [scale=0.38] {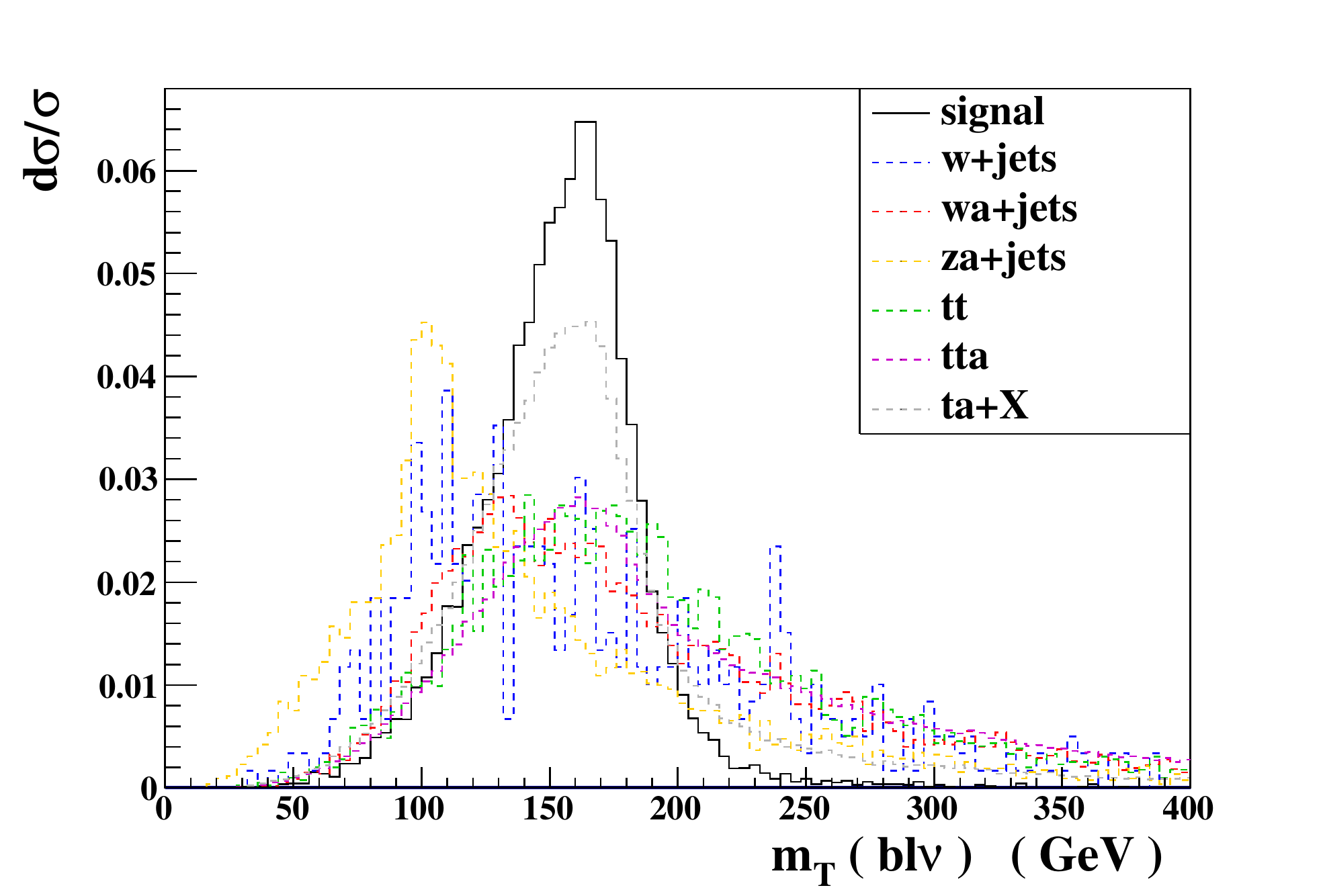}
\vspace{-0.4cm}
\caption{Normalized transverse mass distributions of the $m_{T}(\ell\nu)$ (left) and $m_{T}(\ell\nu b)$ (right) in signal and backgrounds before kinematical cuts at 14 TeV LHC.} \label{Mrlvb}
\end{center}
\end{figure}

In Fig.\ref{Mrlvb}, we plot the transverse mass windows with the preselection cuts by MadAnalysis 5 \cite{ma5}.
The left figure shows the distribution of the transverse mass of the reconstructed $W$, and the right figure for reconstructed top.
The $W$ transverse mass window can reduce $Z\gamma+$jets backgrounds efficiently while affecting the signal slightly. In the right figure, we see that the transverse mass of signal and top processes show narrower peaky distributions, while the multijet backgrounds distribute in a broad transverse mass region as expected. We could use the top transverse mass window to reduce the multijet backgrounds.
From Fig.\ref{Mrlvb}, we require transverse mass cuts as
\begin{eqnarray}
&45{\rm ~GeV}<m_T(\ell \nu)<85{\rm ~GeV},\nonumber\\
&130{\rm ~GeV}<m_T(\ell \nu b)<190{\rm ~GeV}.
\end{eqnarray}

\begin{table}[!htb]
\begin{center}
\caption{\small  The event numbers of the $\ell\nu b\gamma$ signal and backgrounds with $\mathcal{L}=100 $fb$^{-1}$ and $\sqrt{s}=14$ TeV.}
\label{rlvb}
\begin{tabular}{|c|c|c|c|c|c|c|c|}\hline
   &\multicolumn{1}{|c|}{$\ell\nu b\gamma $}
   &\multicolumn{1}{|c|}{  $W$+jets}
   &\multicolumn{1}{|c|}{  $W\gamma$+jets}
   &\multicolumn{1}{|c|}{  $Z\gamma$+jets}
   &\multicolumn{1}{|c|}{  $t\bar{t}$}
   &\multicolumn{1}{|c|}{  $t\bar{t}\gamma$}
   &\multicolumn{1}{|c|}{  $t\gamma+X$} \\
 \hline
preselection cut 1               &$5736.3$    &$18640$   &$19920$   &$2492.9$   & $149752$  & $19973$ & $9481.2$  \\ \hline
preselection cut 2               &$1084.0$    &$2189.3$    &$2378.0$     &  $74.60$      &$38655$     & $5459.5$  & $1683.7$  \\ \hline
$45{\rm ~GeV}<m_T(\ell \nu)<85{\rm GeV} $      &$670.7$    &$1282.3$     &$1471.2$    &   $34.43$     &$15028$     & $2271.6$    &   $852.7$  \\ \hline
$130{\rm ~GeV}<m_T(\ell \nu b)<190$ GeV  &$515.4$    &$375.3$     &$454.5$    &  $11.48$  &  $6696.2$    &$990.8$     & $539.0$     \\ \hline
$S/\sqrt{S+B}$                   &\multicolumn{7}{|c|}{5.265}               \\ \hline
\end{tabular}
\end{center}
\end{table}
We calculate the statistical significance  $S/{\sqrt{(S+B)}}$ for the luminosity of 100 fb$^{-1}$, where $S$ and $B$ denote the number of the signal and background events, respectively. After taking into account transverse mass widow cuts to reject backgrounds, we can further suppress background and gain in the significance up to $\sim5\sigma$, respectively, as represented in Table \ref{rlvb}.

\begin{figure}[!htb]
\begin{center}
\includegraphics [scale=0.38] {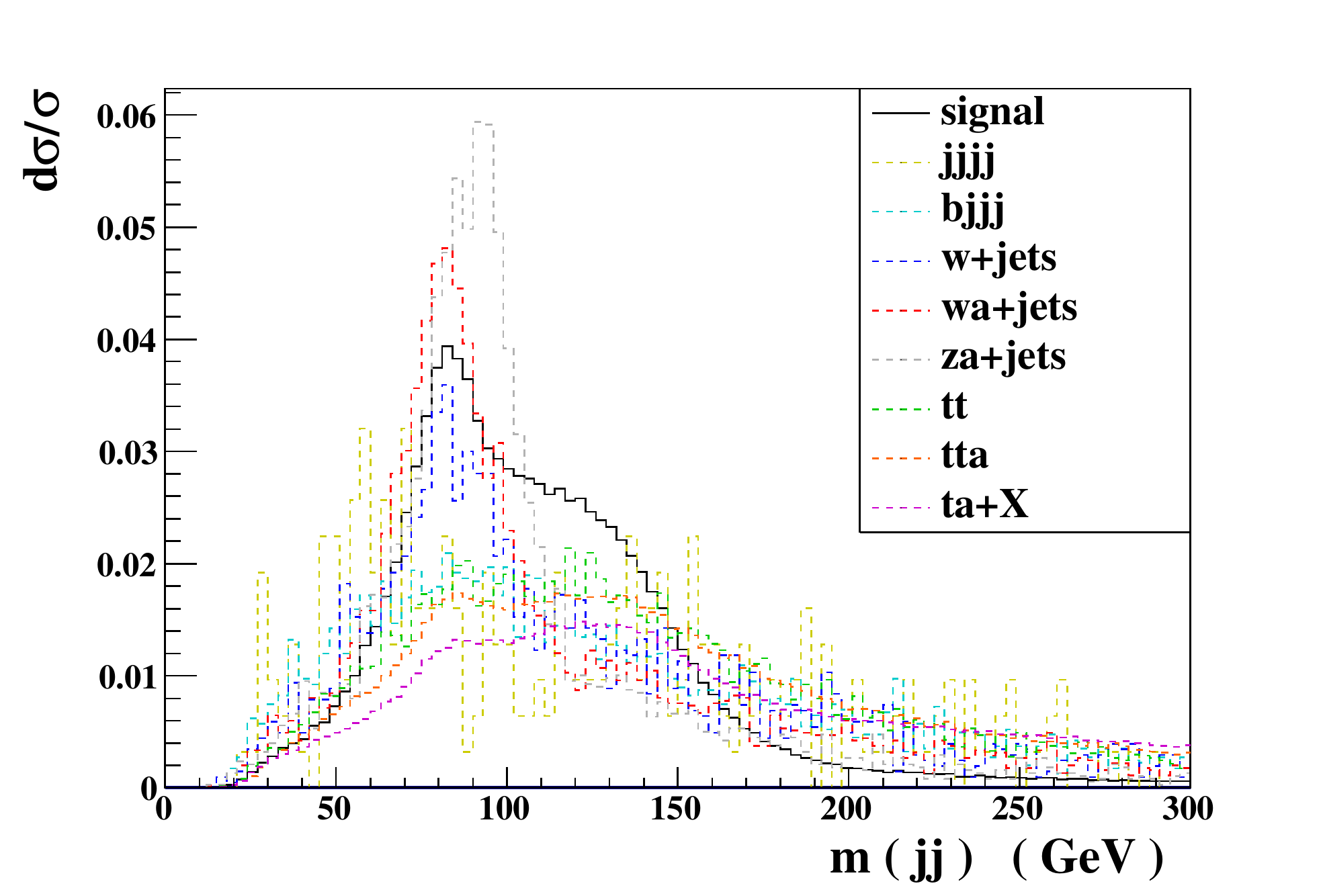}
\includegraphics [scale=0.38] {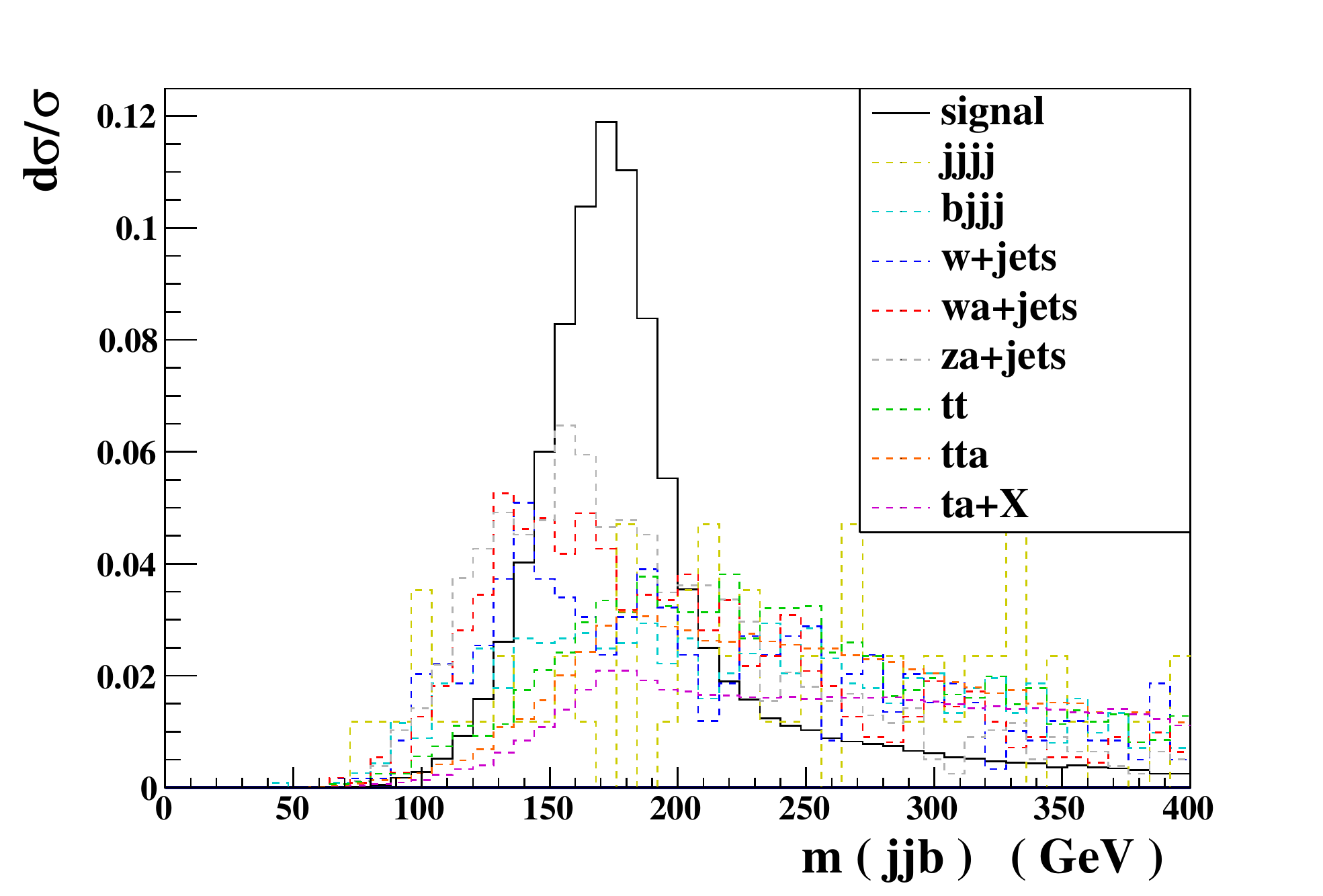}
\vspace{-0.4cm}
\caption{ Normalized invariant mass distributions of the $m(jj)$ (left) and $m(jjb)$ (right) in signal and backgrounds before kinematical cuts at 14 TeV LHC.}
\label{Mrjjb}
\end{center}
\end{figure}

Similar to leptonic mode, we display $m(jj)$ and $m(jjb)$ distributions of $jjb\gamma$ with preselection cuts in Fig.\ref{Mrjjb}.
In the left figure, the sharp peak of signal corresponds to the $W$ mass, while the right part with $m(jj)>100$ GeV is caused by unidentified $b$ quark. Other top background processes represent the same distributions as well. Due to the wide distribution in signal, we claim that the $m(jj)$ cut will not be very effective in improving the significance of the signal. The right figure shows all backgrounds distribute in a broad invariant mass region, while a sharp peaky distribution close to top mass in signal events. As a result, we require the invariant mass $m(jjb)$ to be around the top quark mass window,
\begin{eqnarray}
|m(jjb)-m_t|<35{\rm ~GeV}.
\end{eqnarray}

\begin{table}[!htb]
\begin{center}
\caption{\small  The event numbers of the $jjb\gamma$ signal and backgrounds with $\mathcal{L}=100 $fb$^{-1}$ at 14 TeV LHC.}
\label{rjjb}
\scalebox{0.9}{\begin{tabular}{|c|c|c|c|c|c|c|c|c|c|}\hline
   &\multicolumn{1}{|c|}{Signal $\gamma jjb$}
   &\multicolumn{1}{|c|}{$W$+jets}
   &\multicolumn{1}{|c|}{$W\gamma$+jets}
   &\multicolumn{1}{|c|}{$Z\gamma$+jets}
   &\multicolumn{1}{|c|}{$t\bar{t}$}
   &\multicolumn{1}{|c|}{$t\bar{t}\gamma$}
   &\multicolumn{1}{|c|}{$t\gamma+X$}
   &\multicolumn{1}{|c|}{$4j$ (QCD)}
   &\multicolumn{1}{|c|}{$bjjj$ (QCD)} \\ \hline
Preselection cut 1     & $14071  $ &$ 114996$   &$104115$   & $76843 $   & 167330  &$9418.7$  & $15571$   & 159114715 & 13745400    \\ \hline
Preselection cut 2     & $ 3974.5$ &$  29137$   &$ 15652$   & $11635 $   &$ 63690$ &$3089.3$  & $ 3564.5$ &  42088408 &  2660400    \\ \hline
$|m(jjb)-m_t|<35$ GeV  & $ 1619.7$ &$  4532.5$  &$  2458.4$ & $ 1375.9$  &$ 11261$ &$ 455.9$  & $  271.2$ &   5132732 &   370711    \\ \hline
$S/\sqrt{S+B}$                   &\multicolumn{9}{|c|}{0.6891}               \\ \hline
\end{tabular}}
\end{center}
\end{table}

From Table \ref{rjjb}, we can find that the top mass window cut can reduce about 4/5 background events (the rejected rates even reach $85-90\%$ for 4$j$ and $t\gamma+X$), whilst the signal only loses about 1/2 events.
However, the observability of the $jjb\gamma$ signal with an integrated luminosity of 100 fb$^{-1}$ at 14 TeV LHC is unpromising, with less than $1\sigma$ level statistical significance.

\subsection{The contribution of photon radiation to signal }
\begin{figure}[!htb]
 \vspace{-0.3cm}
 \begin{center}
\includegraphics [scale=0.9] {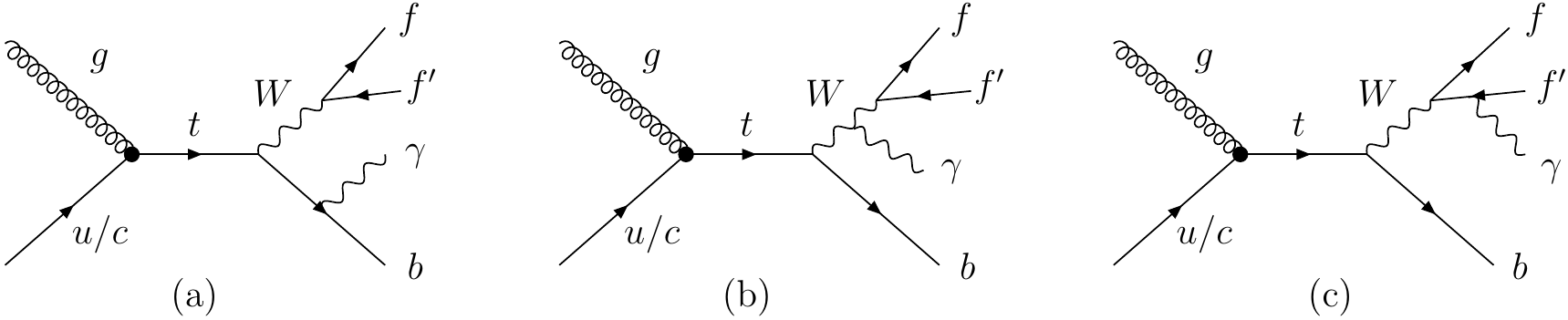}
\caption{Additional Feynman diagrams contributing to the signal of top decay products in association with a photon through $tqg$ vertices. } \label{Feynman2} \vspace{-0.2cm}
\end{center}
\end{figure}
For the signal of anomalous top couplings $tqg$, $t\gamma$ associated production is not the only contribution to $jjb\gamma$ and $\ell\nu b\gamma$ \cite{1003.3173}. The additional Feynman diagrams for $qg \rightarrow f\bar{f}b\gamma$ are depicted in Fig.\ref{Feynman2}. They correspond to direct top production and photon radiation of the $b$ quark, $W$ boson and $W$ decay products, respectively.

\begin{figure}[!htb]
\vspace{-0.5cm}
\begin{center}
\includegraphics [scale=0.25] {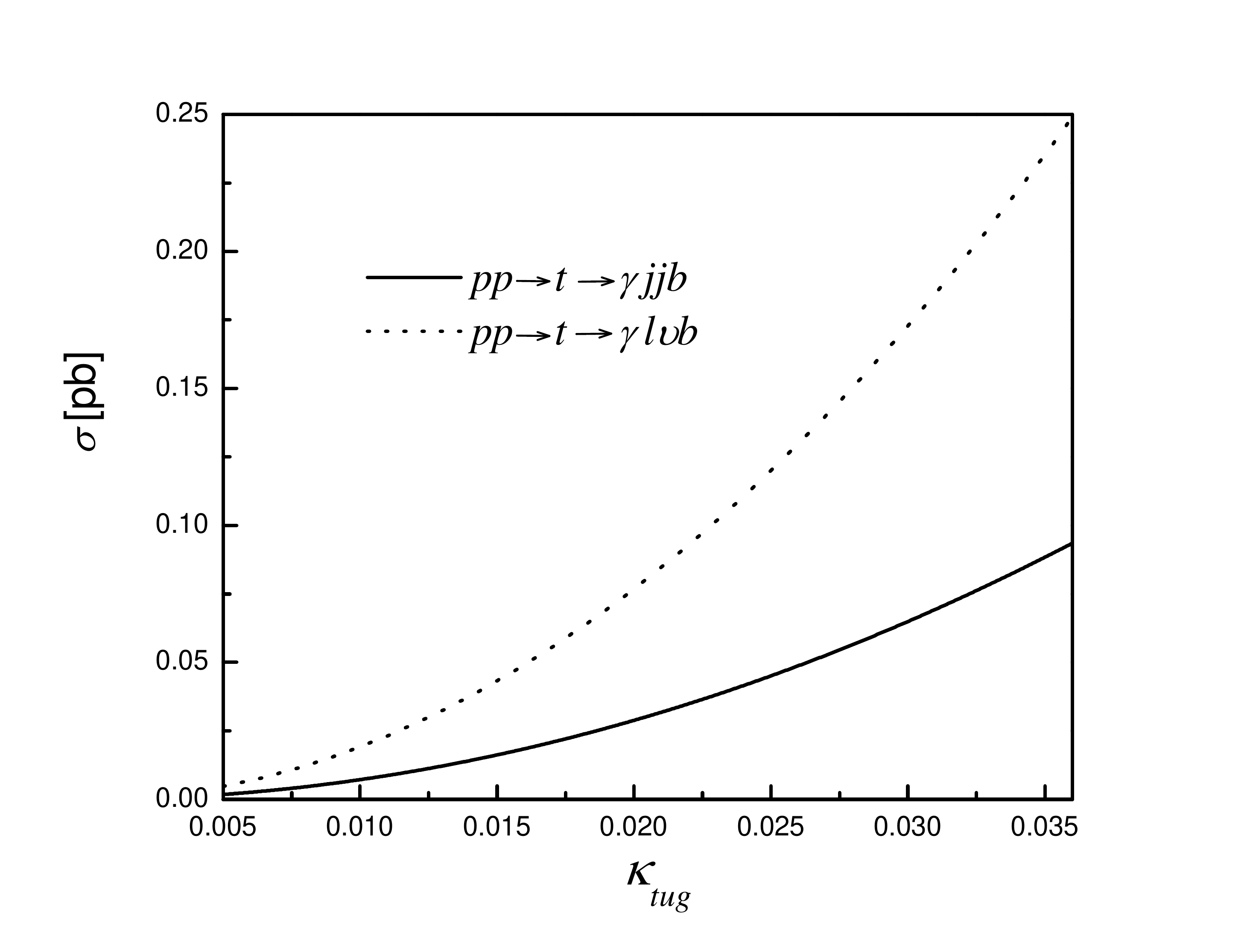}
\includegraphics [scale=0.25] {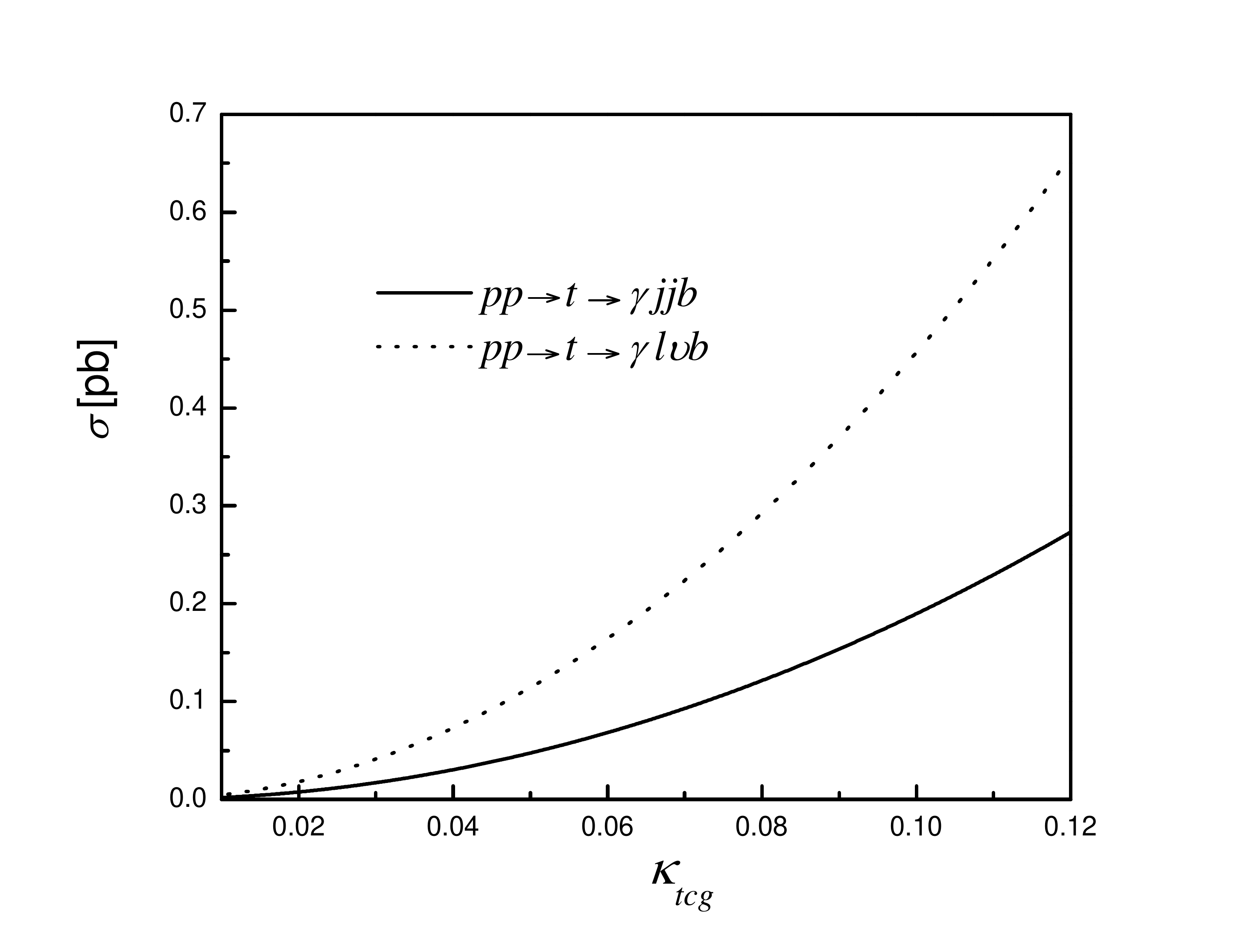}
\caption{ The cross sections of the radiation processes induced by couplings $tug$ (left) and $tcg$ (right) with $p_{T}^\gamma > 15$ GeV.}
\label{radiation cross section}
\end{center}
\end{figure}

We take account of preselection cuts to examine the contributions of the radiation process. By setting basic cuts and $p_T^\gamma > 15$ GeV, we present the cross sections of additional Feynman diagrams with the strengths of anomalous top couplings in Fig.\ref{radiation cross section}.
After setting our preselection cuts, we note that this part of contributions to total cross section have a rapid decrease with the photon transverse momentum increasing. When $p_T^\gamma > 5$ GeV, for $jjb\gamma$ ($\ell\nu b\gamma$), the total cross section which contains the contributions of radiation processes is doubling (tripling) of the $t\gamma$ production with subsequent top decay. When $p_T^\gamma > 40$ GeV, the contributions of additional Feynman diagrams to the cross section quickly decrease to less than $10\%$ of $t\gamma$ production with subsequent top decay.
Thus we conclude that the contributions from $t\gamma$ production with subsequent top decay dominate in part of $p_T^\gamma > 50$ GeV.

\begin{figure}[!htb]
\begin{center}
\includegraphics [scale=0.38] {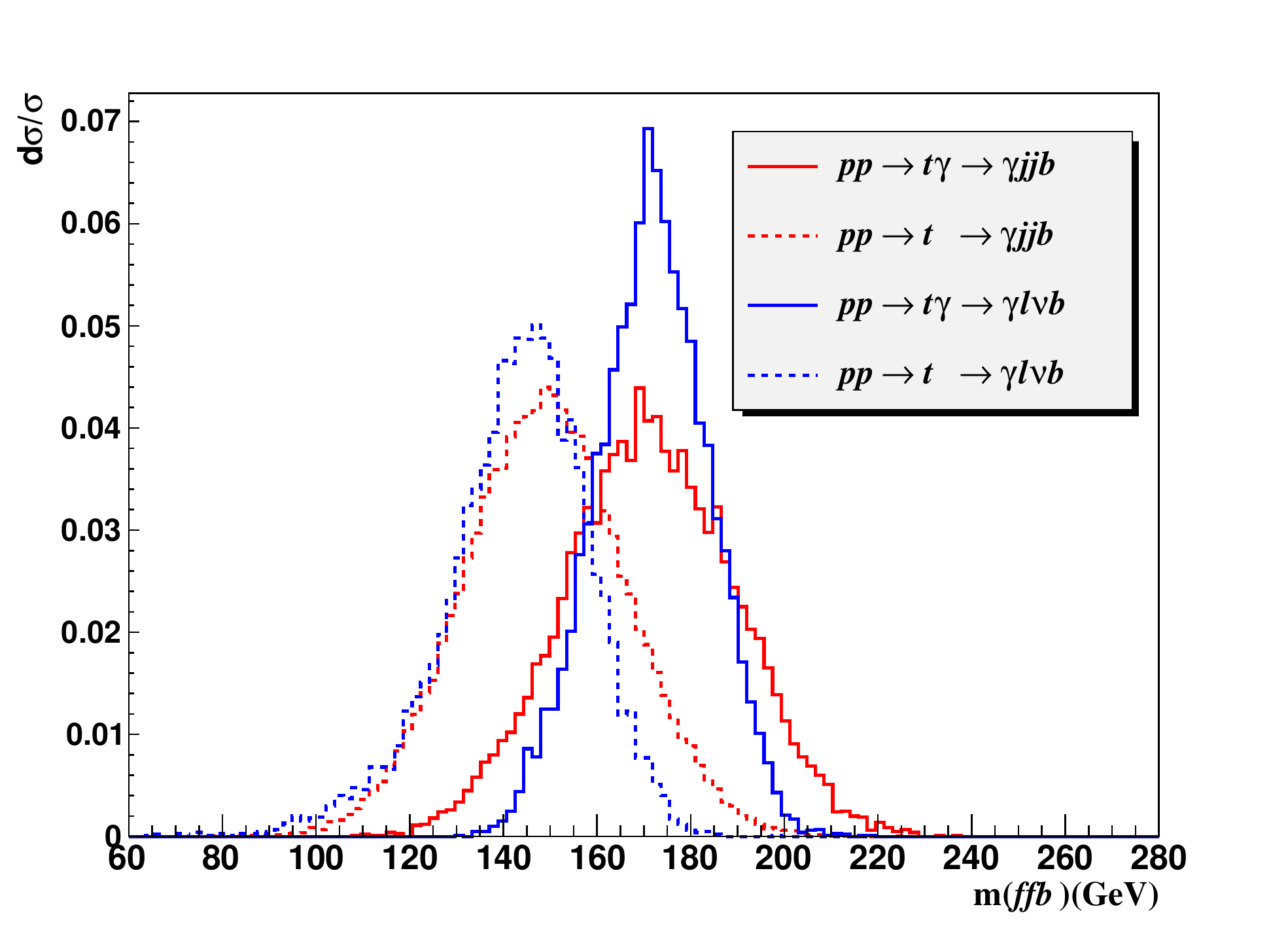}
\includegraphics [scale=0.38] {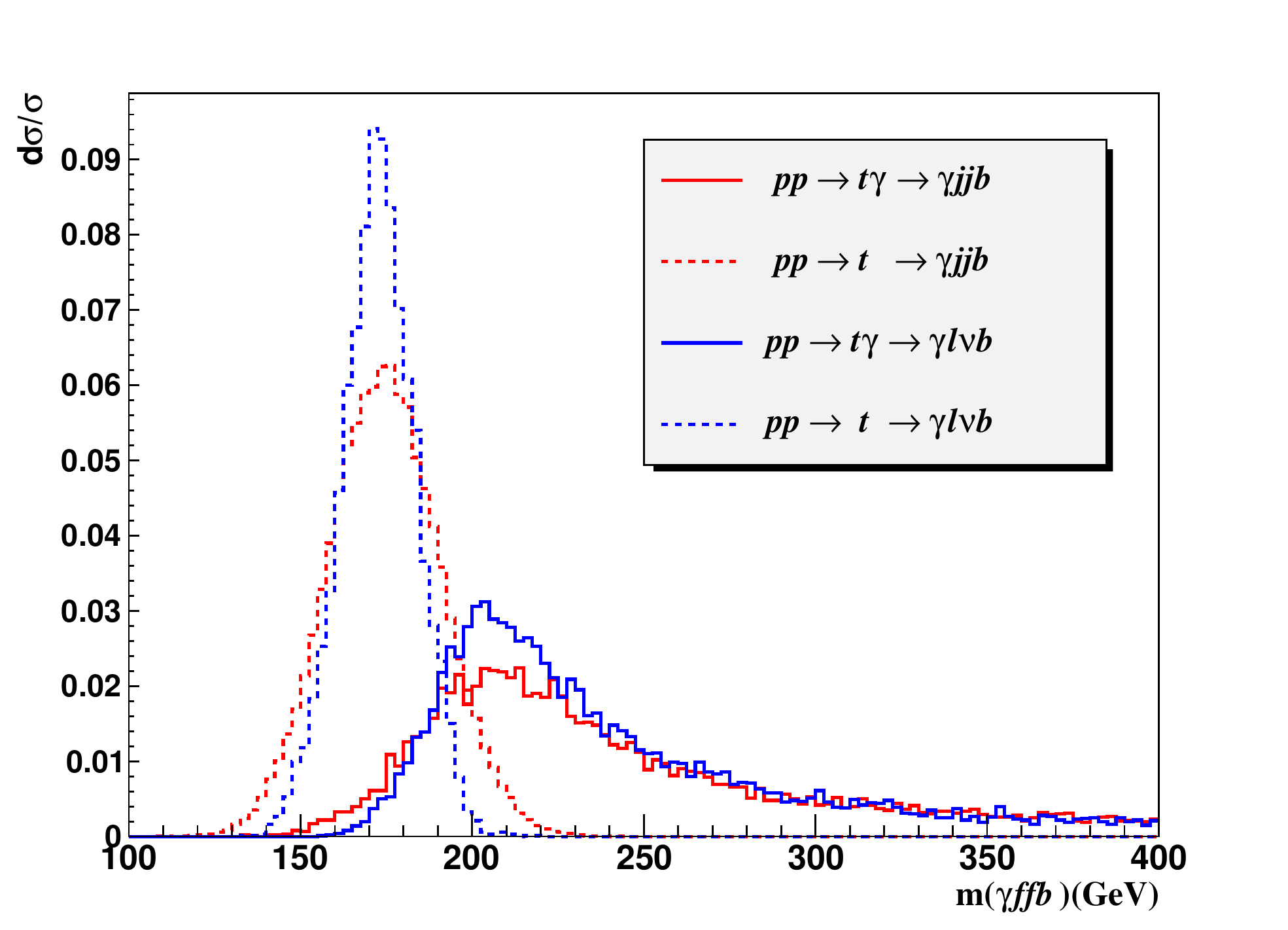}
\caption{ Normalized  distributions of $m(f\bar{f}b)$(left) and $m(f\bar{f}b\gamma)$(right) via different processes with $p_{T}^\gamma > 15$ GeV. The $f\bar{f}$ here denotes $W$ decay products $jj$ or $\ell \nu$.}
\vspace{-0.4cm}
\label{radiation}
\end{center}
\end{figure}

As a probe to research the top quark FCNC, radiation processes could help us to further understand whether it is induced by strong interactions. In this case, it is necessary to consider not only the final states of a top quark plus a photon but also the final state particles reconstructing a top quark. Thus we present the distributions for the $m(f\bar{f}b)$ and $m(f\bar{f}b\gamma)$ in Fig.\ref{radiation}.

The contributions from these additional Feynman diagrams could enhance the signature to help us searching for top FCNC process.
The radiation processes should be considered as a part of the top FCNC signals.

\section{Sensitivity of anomalous top couplings at 14 TeV LHC}
\begin{figure}[!htb]
\begin{center}
\includegraphics [scale=0.25] {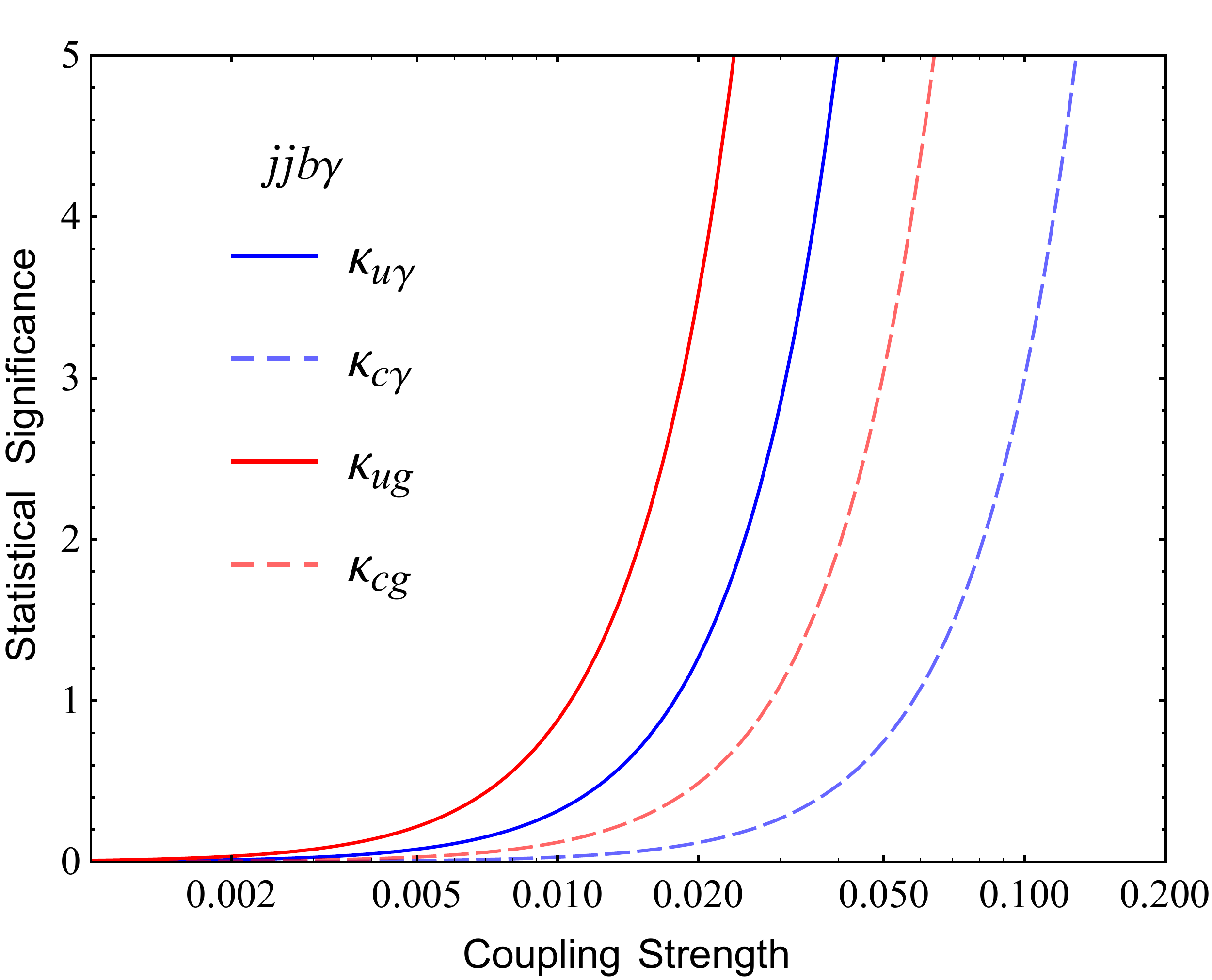}
\includegraphics [scale=0.25] {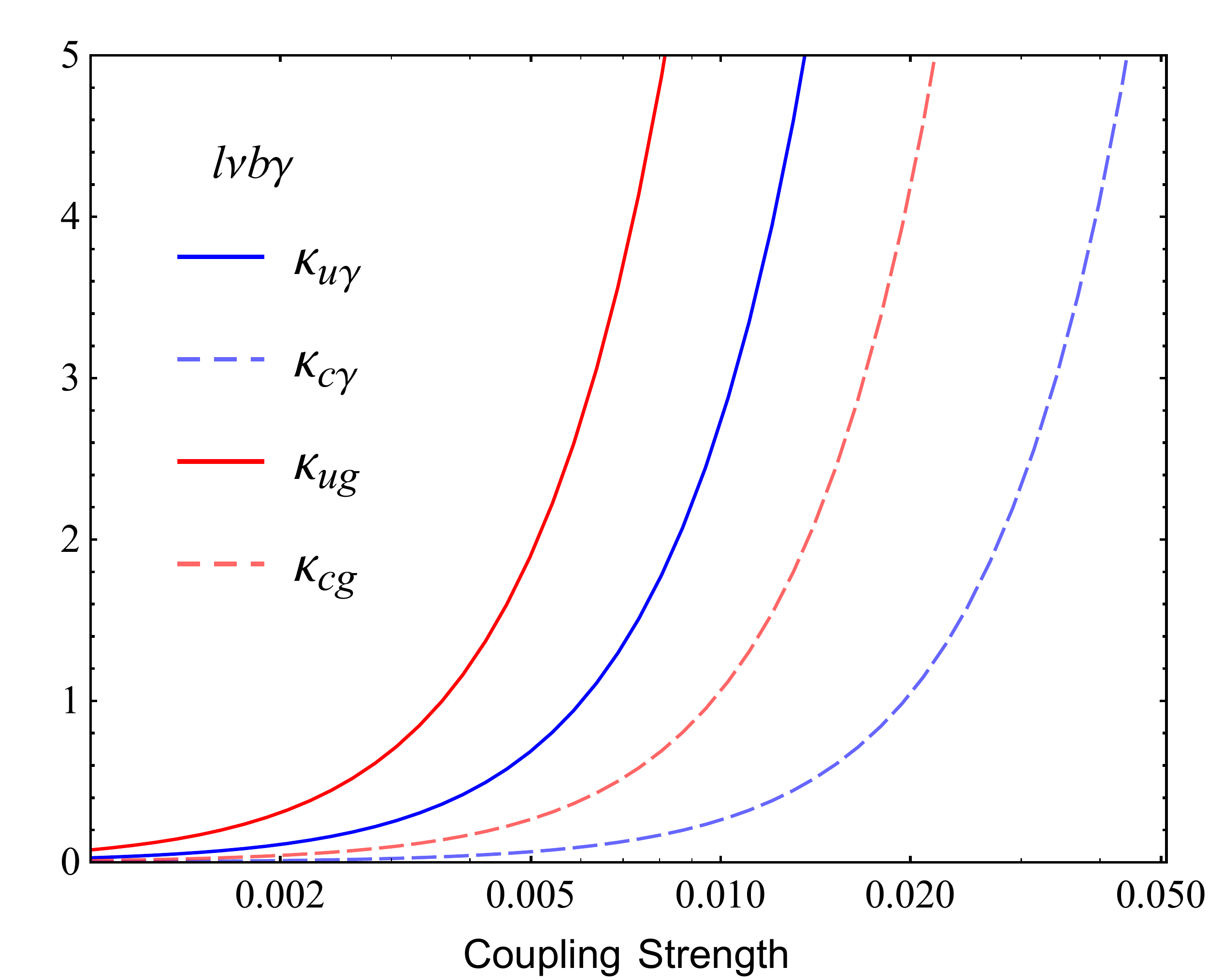}
\vspace{-0.4cm}
\caption{ LHC sensitivity to the considered anomalous top couplings as a function of the coupling strengths after applying the kinematical cuts and event selection.}
\label{fig:SS}
\end{center}
\end{figure}
In this section, we study the sensitivity of anomalous couplings through the $t\gamma$ production at 14 TeV LHC. For an adequate signal modelling, the photon radiation from top quark decay products is taken into account.
The sensitivity of four anomalous top couplings at 14 TeV LHC which is defined as $S/\sqrt{S+B}$ (SS) are presented in Fig.\ref{fig:SS}.
The $SS$ is obtained with the selection strategy talk about in section II.
In the case of assuming a single non-vanishing coupling at a time, four choices of anomalous coupling parameters are subsequently fitted by polynomial functions so that $3\sigma$ and $5\sigma$ discovery ranges are extracted. Assuming that LHC at 14 TeV could collect an integrated luminosity of 100 fb$^{-1}$, we obtain the constraints of one-dimensional discovery limits of anomalous couplings as shown in Table \ref{kappa}.

\begin{table*}[ptbh]
\centering
\caption{$5\sigma$ ($3\sigma$) discovery lower limits on top quark FCNC anomalous couplings.}
\scalebox{0.85}{\begin{tabular}{|c|c|c|c|c|}
\hline
Signal  & $\kappa_{tu\gamma}$ & $\kappa_{c\gamma}$&  $ \kappa_{tug}$  &  $\kappa_{tcg} $  \\
\hline
$\ell\nu b\gamma$ & $0.0136 (0.0105)$ &  $0.0442 (0.0341)$&  $0.0082 (0.0063)$ &  $0.0219 (0.0169)$ \\
$jjb\gamma$      & $0.0398 (0.0308)$ &  $0.1292 (0.1001)$&  $0.0239 (0.0185)$ &  $0.0641 (0.0496)$ \\
\hline
\end{tabular}}
\label{kappa}
\end{table*}
Obviously, the overwhelming QCD multijet backgrounds make the FCNC coupling constants in hadronic mode looser. The $\ell\nu b\gamma$ signal is more sensitive to search for anomalous couplings at 14 TeV LHC.

\begin{figure}[!htb]
\begin{center}
\includegraphics [scale=0.75] {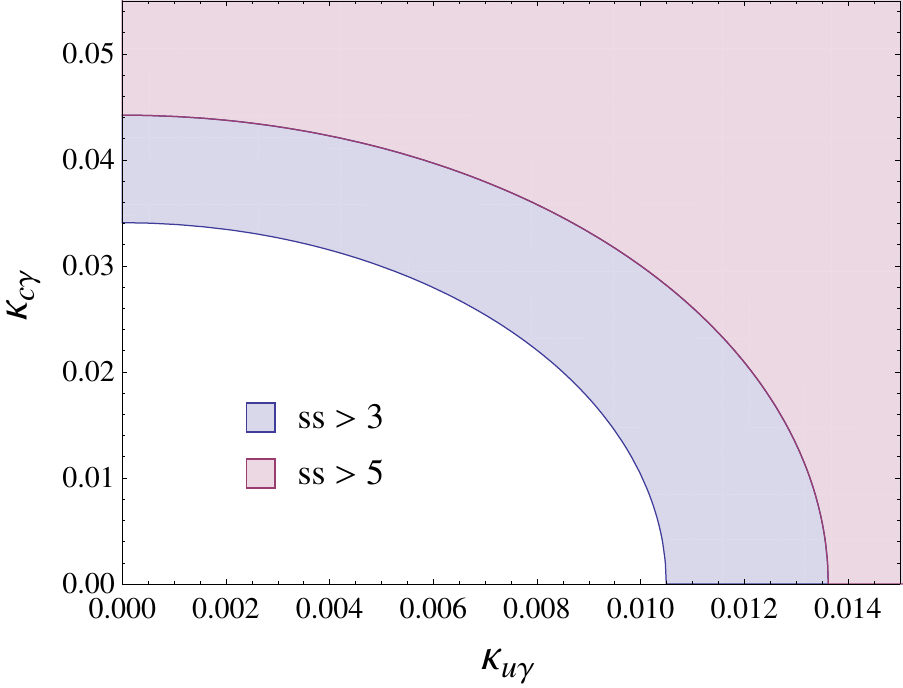}
\includegraphics [scale=0.78] {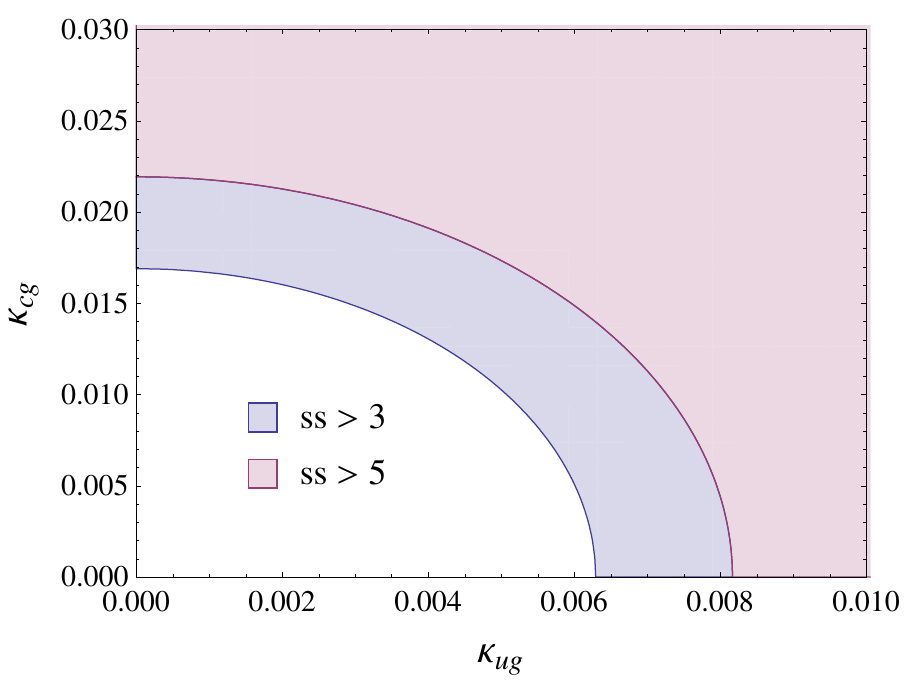}
\caption{3$\sigma$ and $5\sigma$ detection potential regions for the $\ell\nu b\gamma$ signal in weak (left) and strong (right) sector at 14 TeV LHC.} \label{ssrlvb}
\end{center}
\end{figure}

Actually, we allow for a set of non-vanishing couplings simultaneously, either in the weak sector (non-vanishing $\kappa_{tu\gamma}$ and $\kappa_{tc\gamma}$) or in the strong sector (non-vanishing $\kappa_{tug}$ and $\kappa_{tcg}$).
In order to illustrate excluded detection potential regions of anomalous couplings to reach a given statistical significance, we plot the associated $3\sigma$ and $5\sigma$ discovery reaches in $\kappa_{tu\gamma}$-$\kappa_{tc\gamma}$ ($\kappa_{tug}$-$\kappa_{tcg}$) planes for $\ell\nu b\gamma$ at 14 TeV LHC in Fig. \ref{ssrlvb}.
We observe a better sensitivity to flavour-changing interactions with an up quark than with a charm quark, as expected from parton densities, the charm content of the proton being suppressed with respect to its up content.

Since both $tq\gamma$ and $tqg$ operators contribute to the same final state, the interference effects should be considered. If $\kappa_{tu\gamma}=\kappa_{tc\gamma}=\kappa_{tq\gamma}$, $\kappa_{tug}=\kappa_{tcg}=\kappa_{tqg}$, for $p_T^\gamma > 50$ GeV,  the total $t\gamma$ cross section with contributions of $tq\gamma$ and $tqg$ operators is
\begin{equation}
\sigma_{t\gamma} = 158.2\ \big|\kappa_{tq\gamma}|^2 + 457.3\ \big|\kappa_{tqg}|^2 + 153\ \kappa_{tq\gamma}\cdot\kappa_{tqg} \ (\mathrm{pb}) \ .
\end{equation}

\begin{figure}[!htb]
\begin{center}
\includegraphics [scale=1] {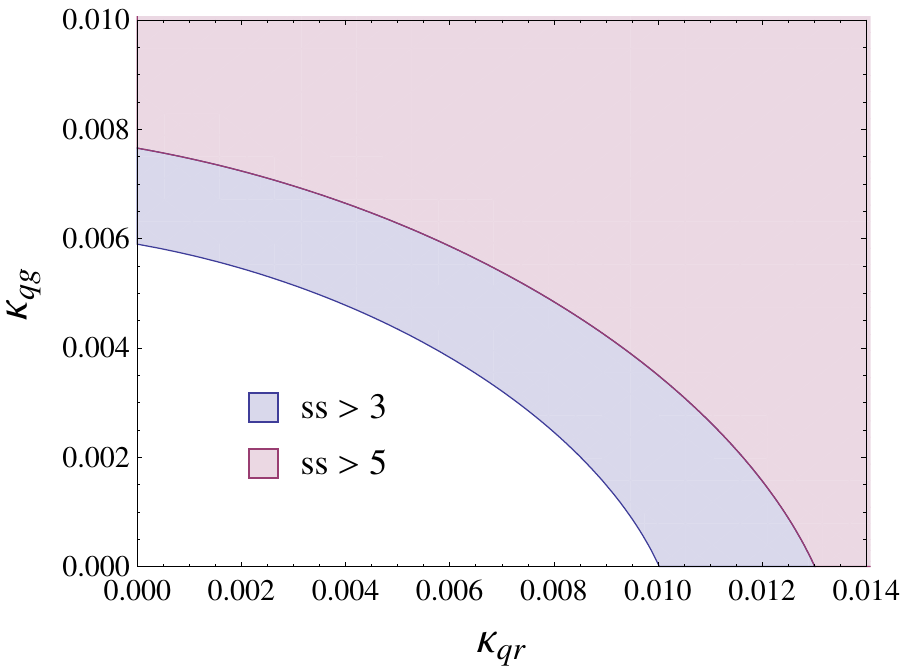}
\caption{3$\sigma$ and $5\sigma$ discovery ranges at 14 TeV LHC in the $\kappa_{tq\gamma}-\kappa_{tqg}$ plane.}\label{tqr-tqg}
\end{center}
\end{figure}

By applying the same selection strategy described as above,
we presented excluded 14 TeV LHC detection potential regions for the $\ell\nu b\gamma$ signal in $tq\gamma$ and $tqg$ plane in Fig.\ref{tqr-tqg}.
Compared to existing researches as regards $tZ$ production and same-sign top quark production\cite{1304.5551,1408.0493},
our results give more sensitive constraints on top anomalous couplings via $t\gamma$ production.

\section{Conclusions }

Many of the extensions of the SM predict that the tree-level FCNC processes could exist.
With Run-II of the LHC, more and more measurements in the top quark sector will be explored with an unprecedented precision. Measurements of single top production allow us to search for deviations from the SM predictions. While these deviations are often interpreted in terms of anomalous top couplings.
The possible deviations can be described by the effects of effective operators, and experimental results can be used to determine useful constraints on each effective operator.
The established deviations can then be evolved up to high scales, and matched to possible new physics scenarios.

In this paper, we have investigated FCNC couplings $tq\gamma$ and $tqg$ in a model-independent way. These interactions lead to possibly significant production rates for $jjb\gamma$ and $\ell\nu b\gamma$ signals via $t\gamma$ associated production with subsequent decay. We have considered the contribution of the single top production with photon radiation off the top decay products to this process. Once $t\gamma$ production processes are eventually discovered, the single top decay with photon radiation processes could help us find out whether or not the origin of the FCNC interactions is in the strong sector. This contribution dominates when $p_T^\gamma$ is small, and it quickly decreases with the photon transverse momentum increasing. For our cuts, the contributions of these processes could enhance the signature, so we include these processes in signal for our analysis of the sensitivity of anomalous couplings.

The sensitivities of 14 TeV LHC to anomalous FCNC couplings $tq\gamma$ and $tqg$ were calculated for both the $\ell\nu b\gamma$ and the $jjb\gamma$ signals. Due to the overwhelming QCD multijet backgrounds, it is challenging to discover the FCNC $t\gamma$ production via the hadronic mode explored at LHC. We found that it is most promising to observe anomalous top couplings via leptonic mode of $t\gamma$ production at the LHC. We further discussed the interference effects on the cross section from contributions of $tq\gamma$ and $tqg$ operators. Then we presented excluded detection potential regions for the $\ell\nu b\gamma$ signal.
With an integrated luminosity of 100 fb$^{-1}$ and $\sqrt{s}=14$ TeV, for a $5\sigma$ discovery, the needed strengths of $tq\gamma$ and $tqg$ couplings are down to magnitude of 0.001-0.01. We hope our results could help search for the signal of anomalous top couplings at 14 TeV LHC in operation.

\section*{Acknowledgement}

\noindent
One of us (Y. C. G.) acknowledge illuminating discussions with Celine Degrande, Cen Zhang, Eric Conte on the analysis of reconstruction at MadGraph School Shanghai 2015. Y. C. G. would like to thank Jian Wang and Zhi-Cheng Liu for valuable discussions. This work was supported in part by the
National Natural Science Foundation of China under Grants Nos. 11275088, 11545012, 11205023 and 11375248, the Natural Science Foundation of the Liaoning Scientific Committee (No. 2014020151) and Liaoning Excellent Talents in University (Grant No. LJQ2014135).

\end{document}